# Mimicing the Kane-Mele type spin orbit interaction by spin-flexual phonon coupling in graphene devices

Zhanbin Bai[1,†], Rui Wang[1,†], Yazhou Zhou[2,†], Tianru Wu[3,†], Jianlei Ge[1], Jing Li[4], Yuyuan Qin[1], Fucong Fei[1], Lu Cao[1], Xuefeng Wang[5], Xinran Wang[5], Shuai Zhang[1,*], Liling Sun[2,*], You Song[4,*], Fengqi Song[1,*]

[1]National Laboratory of Solid State Microstructures, School of Physics and Collaborative Innovation Center of Advanced Microstructures, Nanjing University, Nanjing 210093, China

[2]Institute of Physics, Beijing National Laboratory for Condensed Matter Physics and University of Chinese Academy of Sciences, Beijing 100190, China

[3]State Key Laboratory of Functional Materials for Informatics, Shanghai Institute of Microsystem and Information Technology, Chinese Academy of Sciences, Shanghai 200050, China

[4]Collaborative Innovation Center of Advanced Microstructures, State Key Laboratory of Coordination Chemistry, and School of Chemistry and Chemical Engineering, Nanjing University, Nanjing 210023, China

[5]National Laboratory of Solid State Microstructures, Collaborative Innovation Center of Advanced Microstructures, and School of Electronic Science and Engineering, Nanjing University, Nanjing 210093, China

---

† These authors contributed equally.

* Corresponding authors. S.Z. (szhang@nju.edu.cn), L.S. (llsun@iphy.ac.cn), Y.S. (yousong@nju.edu.cn), F.S. (songfengqi@nju.edu.cn)

**Abstract:**

On the efforts of enhancing the spin orbit interaction (SOI) of graphene for seeking the dissipationless quantum spin Hall devices, unique Kane-Mele type SOI and high mobility samples are desired. However, common external decoration often introduces extrinsic Rashba-type SOI and simultaneous impurity scattering. Here we show, by the EDTA-Dy molecule decorating, the Kane-Mele type SOI is mimicked with even improved carrier mobility. It is evidenced by the suppressed weak localization at equal carrier densities and simultaneous Elliot-Yafet spin relaxation. The extracted spin scattering time is monotonically dependent on the carrier elastic scattering time, where the Elliot-Yafet plot gives the interaction strength of 3.3 meV. Improved quantum Hall plateaus can be even seen after the external operation. This is attributed to the spin-flexural phonon coupling induced by the enhanced graphene ripples, as revealed by the in-plane magnetotransport measurement.

**Key words:**

graphene; spin-orbit interaction; flexural phonon; weak localization

## 1. Introduction

Intense efforts have been made on enhancing the spin orbit interaction (SOI) of graphene because of the potential quantum spin Hall (QSH) effect [1-15], which was firstly proposed on graphene and recently demonstrated in HgTe/CdTe quantum wells [4, 7], InAs/GaSb [10, 12], single-layer FeSe [14] and monolayer $WTe_2$ [15]. The critical obstacle for the graphene's QSH is that the chemical potential has to be tuned into the SOI-generated gap of the Dirac cone. However, the SOI gap is very small for pristine graphene, i.e. ~$10^{-5}$ to ~$10^{-6}$ eV [16, 17]. The SOI-enhancing proposals include functionalizing the graphene with some light atoms, e.g. hydrogenated graphene [18, 19] and fluorinated graphene [20], decorating the graphene with metallic atoms, such as In [21], Au [22, 23], and attaching graphene to strong spin-orbit coupling materials, such as $Bi_2Te_2Se$ [24], $Bi_2Se_3$ [25] and have been implemented with the improved SOI of ~$10^{-2}$ eV [8, 18, 20-28].

However, in order to achieve the QSH, the SOI type has to be the Kane-Mele (KM) desired type with the Hamiltonian of $\sim\tau_z\sigma_z s_z$, because the common Rashba SOI ($\sim\vec{z}\cdot(\vec{s}\times\vec{\sigma})$) may break the QSH transport by mixing different branches of spin channels [16]. This is reasonable since the KM type SOI is originated from the symmetry of the graphene honeycomb lattice. Furthermore, the above external decoration may simultaneously suppress the mobility of graphene and subsequently destroy rather fragile QSH states.

Here we show that the KM type SOI can be mimicked by a spin-flexual phonon coupling picture, which is implemented by decorating the graphene devices using $Na[Dy(EDTA)(H_2O)_3]\cdot 5H_2O$ (EDTA-Dy) molecules. The Elliot-Yafet (EY) plot of spin scattering time and carrier elastic scattering time gives the SOI strength of 3.3 meV. The graphene devices even present improved half-integer quantum Hall transport after decorated.

## 2. Experimental

The graphene devices (figure 1(a)) were fabricated on a single crystal grain of chemical vapor deposition grown graphene sheet, which were transferred onto the Si substrate with a 300-nm-thick $SiO_2$ layer. Standard e-beam lithography and e-beam evaporation was used to make the Hall bar devices. The EDTA-Dy complex is prepared according the improved methods [29] (see Supporting Materials, Note I).

The influence of EDTA-Dy coating on graphene was investigated. Transport measurements were done in Cryomagnetics' C-Mag system with the low-frequency lock-in technique. Pristine graphene was measured in the system and then it was processed by a little drop of EDTA-Dy solution. The processed graphene was kept in fuming cupboard for several minutes to dry off. EDTA-Dy decorated graphene was then recooled and measured in the system. A vector magnet is used in the in-plane field measurement.

## 3. Result and discussion

### 3.1. Improved transport after the decoration of EDTA-Dy

Figure 1(b) shows the Raman spectrum of EDTA-Dy decorated graphene, where the 2D/G peak ratio is 1.44 and the full width half maximum (FWHM) of the 2D peak is 25.01 $cm^{-1}$. These confirm the monolayer thickness of the graphene. Transport parameters of the four graphene devices are listed in Table I.

Figure 1(c) shows the back gate voltage ($V_g$) dependent longitudinal resistance ($R_{xx}$) of the EDTA-Dy decorated graphene at 2 K, 20 K and 290 K, and the bipolar behavior is obvious. The density of hole carriers decreases after the EDTA-Dy decoration. It means the electronic transfer from EDTA-Dy to graphene, which can be understood through the structure of EDTA-Dy. The EDTA ligand in EDTA-Dy complex has four $–COO^-$ groups which chelate with one $Dy^{III}$ ion. The $–COO^-$ group has free $O^{2-}$ with large electron density. The $O^{2-}$ in $–COO^-$ group has a strong tendency to interact with the π electrons of the benzene ring of the graphene.

Interestingly, the performances of the graphene devices are improved after the EDTA-Dy decoration, which is also seen in graphene grafted with Pt-porphyrins [30]. The carrier mobility unexpectedly changes from 1919 $cm^2V^{-1}s^{-1}$ to 3226 $cm^2V^{-1}s^{-1}$ after decoration. And the quantum Hall effect (QHE) is observed at 12 T and 2 K, as shown in figure 1(d), with the Hall conductivity $\sigma_{xy}$ goes quantized and the longitudinal resistivity $\rho_{xx}$ approaches zero. The 4($n$ + 1/2) plateau series is the characteristic half-integer QHE of monolayer graphene [31, 32].

### 3.2. Suppressed WL and selectively-enhanced KM-type SOI

The SOI alters the interference in a pair of time-reversal loops, which causes the phase shift and quantum modification to the low-field magnetoresistance (MR) [33, 34]. For graphene, E. McCann's weak localization (WL) theory describes the low-field WL correction with a KM-type SOI to the magnetoconductivity (MC) as below [35, 36]

$$\Delta\sigma(\mathrm{B}) = -\frac{e^2}{2\pi h}\left[-2F\left(\frac{B}{B_\varphi + B_{KM}}\right) + 4F\left(\frac{B}{B_\varphi + B_* + B_{KM}}\right) + 2F\left(\frac{B}{B_\varphi + 2B_i + B_{KM}}\right)\right]$$

$$F(\mathrm{z}) = \ln z + \psi\left(\frac{1}{2} + \frac{1}{z}\right), B_{\varphi,i,*,KM} = \frac{\hbar c}{4De}\tau_{\varphi,i,*,KM}^{-1} \quad (1)$$

where $\psi$ is the digamma function, $D$ is the diffusion coefficient, $\tau_\varphi^{-1}$ is the inelastic dephasing rate, $\tau_*^{-1}$ is the intravalley symmetry-breaking scattering rate, $\tau_i^{-1}$ is the intervalley scattering rate, and $\tau_{KM}^{-1}$ is the intrinsic KM spin relaxation rate. The relaxation length scales are related to the relaxation times as $L_{\varphi,\mathrm{i},*} = \sqrt{D\tau_{\varphi,\mathrm{i},*}}$. E. McCann *et al* further predicts that the type of SOI can be identified by analyzing the WL features and spin relaxation mechanism [35, 37]. When intense SOI is introduced into graphene, weak antilocalization (WAL) and D'yakonov-Perel (DP) spin relaxation suggests the dominance of the Rashba-type SOI, while the suppressed WL and EY spin relaxation indicates the dominance of KM-type SOI.

The low-field MC ($\sigma_{xx}$) of the EDTA-Dy decorated graphene with gate voltage is shown in figure 2(a). The MC is sharply suppressed at zero fields, showing the suppressed WL, which agrees with the KM SOI dominant case [35]. The transport parameters can be extracted from fitting with Eq. (1), including the spin scattering time $\tau_{KM}$. The dephasing length $L_\varphi$, intervalley scattering length $L_i$ and intravalley scattering length $L_*$ is 650 nm, 60 nm and 20 nm respectively. The $\tau_{KM}$ against $V_g$ at different temperatures are plotted in figure 2(b), and $\tau_{KM}$ increases with $V_g$ away from the charge neutral point (~62 V). It increases to over 400 ps by two orders at the negative gate voltages. This demonstrates the convenient spin transport control in EDTA-Dy decorated graphene.

Further discussions reveal the spin relaxation mechanism in the devices. The DP procedure occurs through spin precessions while the EY procedure develops through the spin flips by the scatters. They cause distinct dependence of spin relaxation time on the elastic scattering time $\tau_p$, where $\tau_p = h\sigma/(2e^2 v_F\sqrt{\pi n})$. From the $\sigma_{xx}$ and carrier density at each $V_g$, $\tau_p$ can be obtained. We find that $\tau_{KM}$ increases monotonically with $\tau_p$, as shown in figure 2(c), which is the signature of the EY spin relaxation. We make the EY plot $\tau_{KM} = \tau_p(\varepsilon_F/\Delta_{KM})^2$ in figure 2(c), where $\Delta_{KM}$ is the strength of the SOI causing the spin relaxation and $\varepsilon_F$ is the chemical potential. The nice fittings indicate the dominance of the EY relaxation. Thus, the effective SOI strength of the EDTA-Dy decorated graphene can be obtained.

The SOI strength increases with rising temperature both in pristine and decorated graphene as shown in figure 2(d). It is believed that the SOI ($\Delta_{D0}$) which generates the Dirac cone gap will not change with temperature. We therefore extrapolate the data along the linear fitting and expect $\Delta_{D0}$ from the intercepts at zero temperature. It is found that $\Delta_{D0}$ is quite small for pristine graphene, while it achieves 3.3 meV after the EDTA-Dy decoration. Such $\Delta_{D0}$ has been large enough to accommodate the QSH states. The temperature dependent SOI is attributed to the electronic scattering by the impurity centers [38, 39].

### 3.3. Decoration induced ripples revealed by in-plane magnetic field

As stated above, both the suppressed WL and successful EY plot reveals the selectively enhanced KM-type SOI while decorating the graphene by EDTA-Dy. Such decoration should induce more scattering, but the present devices even show improved quantum transport, which forms the main question of this work. We propose the ripples caused by the stretching EDTA-Dy coating might contribute to the pseudo magnetic field (gauge field), which prefers the KM-type SOI in graphene.

Such ripples are revealed by the atomic force microscopy (AFM) as shown in figure

3(a). The root-mean-square (RMS) roughness of pristine graphene is 0.66 nm. After EDTA-Dy decoration, the RMS roughness increases to 1.64 nm. Similar ripples have been described in PTSA coated graphene [40]. It can be understood by the interfacial stretching due to the different temperature dependent shrinking of graphene and the organic film. Such ripples even detach some part of graphene from the substrates and suppress the charge scattering from the substrates [41, 42].

To obtain the in-situ RMS information of the decorated graphene (avoid the artificial effect while removing the coating layer), we measure the MR of the device under in-plane fields. As proposed by J. A. Folk *et al*, the ripple configuration in graphene can be measured by applying an in-plane magnetic field, when a random vector potential (RVP) field is generated and modifies the MR [43]. The RVP-modified $\Delta\rho(B_{\|})$ for Gaussian ripples can be calculated by a Boltzmann approach [43]:

$$\Delta\rho(n,\theta,B_{\|}) = \frac{\sin^2\theta+3\cos^2\theta}{4}\frac{1}{\hbar|n|^{3/2}}\frac{Z^2}{R}B_{\|}^2 \quad (2)$$

where $\theta$ is the angle between the current flow and $B_{\|}$, as shown in the inset of figure 3(b). $Z$ is RMS height and $R$ is correlation length. The graphene before and after EDTA-Dy decoration is measured with $\theta \approx 10^\circ$ and $80^\circ$, which gives $Z^2/R \approx 0.14$ nm and 0.67 nm respectively by fitting Eq. (2).

Further measurements have to be carried out to obtain the RMS ($Z$). Parallel-field -generated magnetic flux through the ripples causes orbital effect, due to which the phase-coherent WL can be suppressed. For this, we design the experiment to measure the WL features driven by a perpendicular field while applying a series of fixed parallel fields.

The WL features driven by a perpendicular field can be formulated as below [44]

$$\Delta\sigma(B_{\perp}) = \frac{e^2}{\pi h}\left[F\left(\frac{\tau_B^{-1}}{\tau_\varphi^{-1}}\right) - F\left(\frac{\tau_B^{-1}}{\tau_\varphi^{-1}+2\tau_i^{-1}}\right) - 2F\left(\frac{\tau_B^{-1}}{\tau_\varphi^{-1}+\tau_i^{-1}+\tau_*^{-1}}\right)\right] \quad (3)$$

where $\tau_B^{-1} = 4eDB_{\perp}/\hbar c$. In figure 3(c), the WL scattering rates are extracted by fitting with Eq. (3), in which $\tau_\varphi^{-1}, \tau_i^{-1}$ and $\tau_*^{-1}$ are $1.14\times10^{11}$ $s^{-1}$, $6.01\times10^{10}$ $s^{-1}$ and $1.18\times10^{14}$ $s^{-1}$ respectively for $B_{\|} = 0$.

The WL dip is further suppressed after applying an in-plane field, as shown in figures 3(c) and 3(d), where Eq. (3) is fitted for a series of $B_{\|}$. During the fitting, $\tau_i^{-1}$ and $\tau_*^{-1}$ are fixed. Extracted values of $\tau_\varphi^{-1}$ plotted against $B_\|^2$ are shown in figure 3(e), and linear fitting confirms the dependence in Eq. (4) proposed by J. A. Folk *et al* [43], with $Z^2R = 1.07$ nm$^3$.

$$\tau_\varphi^{-1} \to \tau_\varphi^{-1} + \sqrt{\pi}(e^2/\hbar^2)vZ^2RB_\|^2 \quad (4)$$

From the values of $Z^2/R$ and $Z^2R$, $Z$ = 0.62 nm and $R$ = 2.7 nm are obtained for Gaussian-correlated ripples for the pristine graphene. Similarly, $Z$ = 1.48 nm and $R$ = 3.2 nm are the ripple dimensions of EDTA-Dy decorated graphene. This confirms the ripple generation by the EDTA-Dy coating.

**3.4. Spin-flexural phonon coupling mode in rippled graphene**

Here we propose that the EY spin relaxation can be interpreted by the flexural phonon mode in rippled graphene. This is partially based on the previous study on all symmetry-adapted spin-phonon couplings [45], where the phonons vibration associated to the $B_2$ irreducible representation generates a correction term to the electrons, $H_{ph} = g\sigma^z s^z h(\boldsymbol{r})^2$, where $h(\boldsymbol{r})$ is the vertical displacement of the atoms, $\sigma$ is Pauli matrices defined in the sublattice, and $s$ is spin. Considering the thermal average, the phonon mode further results in a local term that reads as $H_{KM} = \Delta(\boldsymbol{r})\sigma^z s^z$. $\Delta(\boldsymbol{r})$ is proportional to $\langle h^2 \rangle$. The low-energy effective model reads as

$$H_{eff} = -i\hbar v_F \sigma \cdot \nabla + \Delta(r)\sigma^z s^z \quad (5)$$

This effective Hamiltonian becomes a KM-like term and explains the dominance of the KM-type SOI in our study.

This model reasonably predicts that only the in-plane spins would undergo the spin-flip during scattering since the $z$ component of spins are conserved. $\Delta(r)$ can be treated as disorder characterized by the Gaussian type delta function-correlated function, i.e., $\langle \Delta(r) \rangle = 0$ and $\langle \Delta(r)\Delta(r') \rangle = \bar{\Delta}\delta(r - r')$. Then dealing with the disorder average [46-48] (see supplementary), it can be shown that a global effective KM-like term would take place in mean-field level. Solving the corresponding Schrödinger equation, one can obtain the spin-flip probability which reads as $\alpha = \frac{\bar{\Delta}^2}{4\epsilon_F^2}$.

This finally leads to a EY-type spin relaxation time that satisfy $\tau_s = \varepsilon_F^2 \tau_p / \bar{\Delta}^2$, accounting for the results in figure 2.

To further study the temperature dependent spin lifetime, we resort to a second quantization form of the phonon field. In this form, the spin-phonon interaction is reduced to

$$H_{ph} = \sum_{\boldsymbol{k},\boldsymbol{q},\boldsymbol{q}'} V_{\boldsymbol{k},\boldsymbol{q},\boldsymbol{q}'} C^{+}_{\boldsymbol{k}+\boldsymbol{q}+\boldsymbol{q}'} \sigma_z s_z C_{\boldsymbol{k}} (d_{\boldsymbol{q}} + d^{+}_{-\boldsymbol{q}})(d_{\boldsymbol{q}'} + d^{+}_{-\boldsymbol{q}'}) \qquad (6)$$

where $V_{\boldsymbol{k},\boldsymbol{q},\boldsymbol{q}'} = \frac{g\hbar}{2\rho\sqrt{\omega_{\boldsymbol{q}}\omega_{\boldsymbol{q}'}}}$, $\omega_{\mathbf{q}}$ is the phonon dispersion, $\rho$ is the mass density of the Carbon atoms, and $C_{\boldsymbol{k}}$ is a four-dimensional electron spinor in sublattice and spin space. Evaluating the spin-flip scattering probability and using the Fermi's golden sum rule, the spin relaxation rate can be obtained. For higher temperature region with $T > \hbar\omega_{\boldsymbol{F}}/k_B$, which is the experimental case [49], we arrive at

$$\frac{1}{\tau_s} = AT^2 + BT + C \qquad (7)$$

where *A* and *B* are the second order and first order *T* dependent coefficient respectively, while *C* is a constant representing the zero temperature residue contribution [45]. As an estimation, neglecting the dispersion of phonon and replacing the integral by an averaged frequency $\bar{\omega}$, then *A* and *B* can be found to satisfy $A = \frac{64 g^2 k_B^2 k_F^3}{\pi \hbar^2 v_F \rho^2 \bar{\omega}^4}$ and $B = \frac{64 g^2 k_B k_F^3}{\pi \hbar v_F \rho^2 \bar{\omega}^3}$. For experimental values, we take $\rho = 7.6 \times 10^{-7}$ kg·m$^{-2}$ and $\bar{\omega}$ = 10 meV. In addition, *g* is related to the strength of KM-like SOC. From figure 2(d), *g* can be set to be $g \sim 2$ eV·Å$^{-2}$. Then the parameters can be estimated to be $A \sim 2.0 \times 10^{-6}$ K$^{-2}$ps$^{-1}$ and $B \sim 2.3 \times 10^{-4}$ K$^{-1}$ps$^{-1}$.

In figure 4, we fit the experimental $\frac{1}{\tau_s} - T$ curve according to Eq. (7), the fitted parameters of the first graphene device after EDTA-Dy decoration are found to be $A \sim 1.9 \times 10^{-6}$ K$^{-2}$ps$^{-1}$ and $B \sim 1.86 \times 10^{-4}$ K$^{-1}$ps$^{-1}$, in well agreement with our estimated values. As can be seen from figure 4, while it is almost negligible in the pristine sample, an obvious temperature dependent spin life time is found to emerge in the decorated graphene.

## 4. Conclusion

We can selectively enhance the KM-type SOI in graphene by the EDTA-Dy decoration, as evidenced by the suppressed WL and simultaneous EY spin relaxation. The SOI strength of 3.3 meV is achieved. The QHE can be seen with a $4(n + 1/2)$ series of plateaus even after decoration. We believe the interfacial stretching and local ripple configuration may achieve the simultaneous external SOI control and transport improvement. A spin-flexural phonon coupling model is proposed. Such subtle control of the electronic properties paves the way towards the topological insulator states in graphene.

## Acknowledgments

We gratefully acknowledge the financial support of the National Key R&D Program of China (No. 2017YFA0303203), the National Natural Science Foundation of China (Grant Nos: U1732273, U1732159, 12025404, 11904166, 11904165, 61822403, 11874203, 11834006, 91622115, 11522432, 11574217 and 21571097), the Natural Science Foundation of Jiangsu Province (Grant BK20190286), the PAPD project, and the Fundamental Research Funds for the Central Universities.

**Figures and Tables**

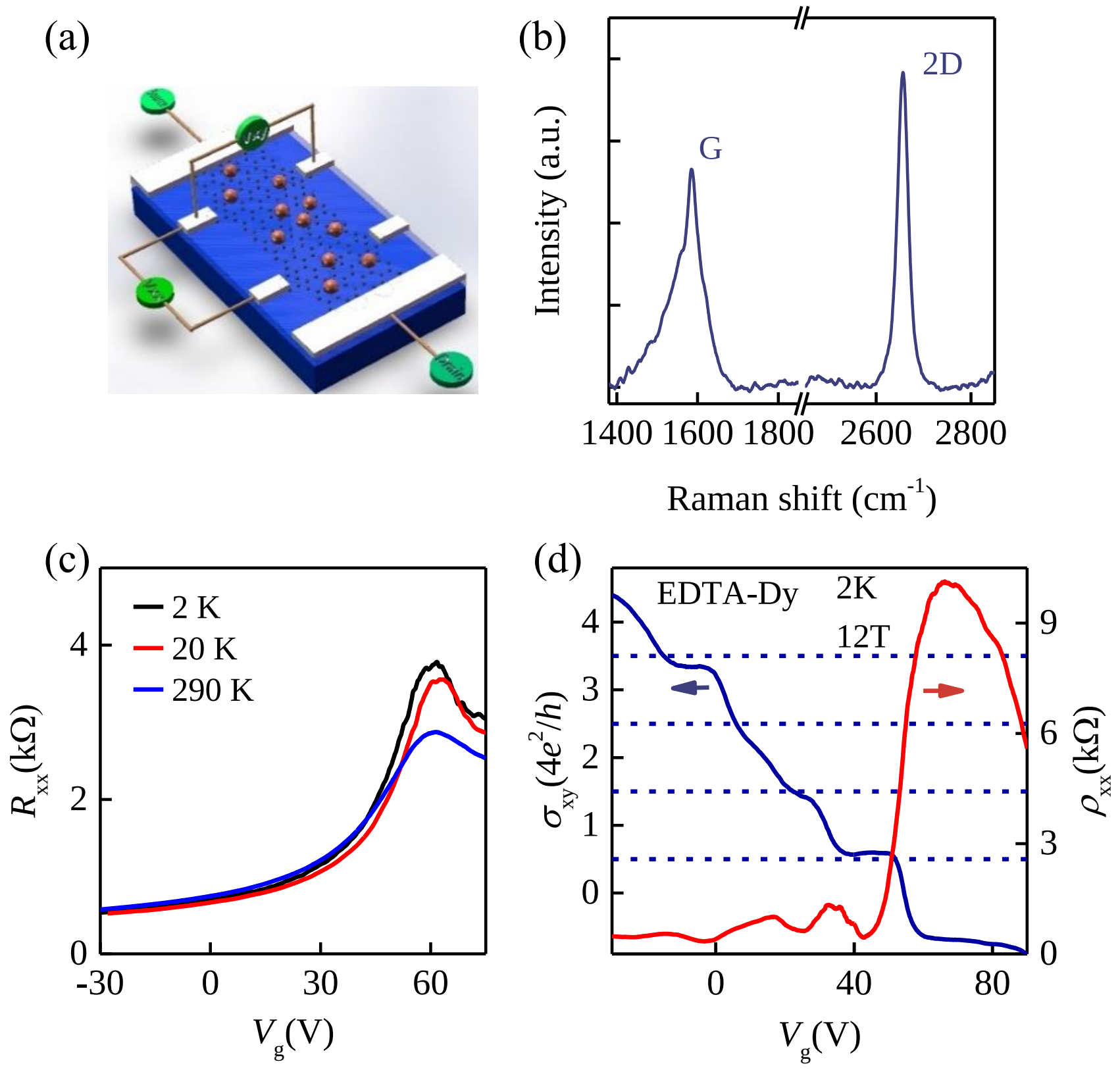

**Figure 1. The EDTA-Dy decorated graphene and its transport.** (a) The schematic configuration of the graphene device, where the EDTA-Dy (orange ball) coats the graphene. (b) Raman spectrum of EDTA-Dy decorated graphene, indicating the sample is monolayer graphene. (c) Resistance as a function of back gate voltage ($V_g$) for EDTA-Dy decorated graphene at 2, 20 and 290 K. (d) $V_g$ dependent longitudinal resistivity $\rho_{xx}$ and Hall conductivity $\sigma_{xy}$ at 12 T and 2 K, where $\sigma_{xy}$ goes quantized when $\rho_{xx}$ approaches zero.

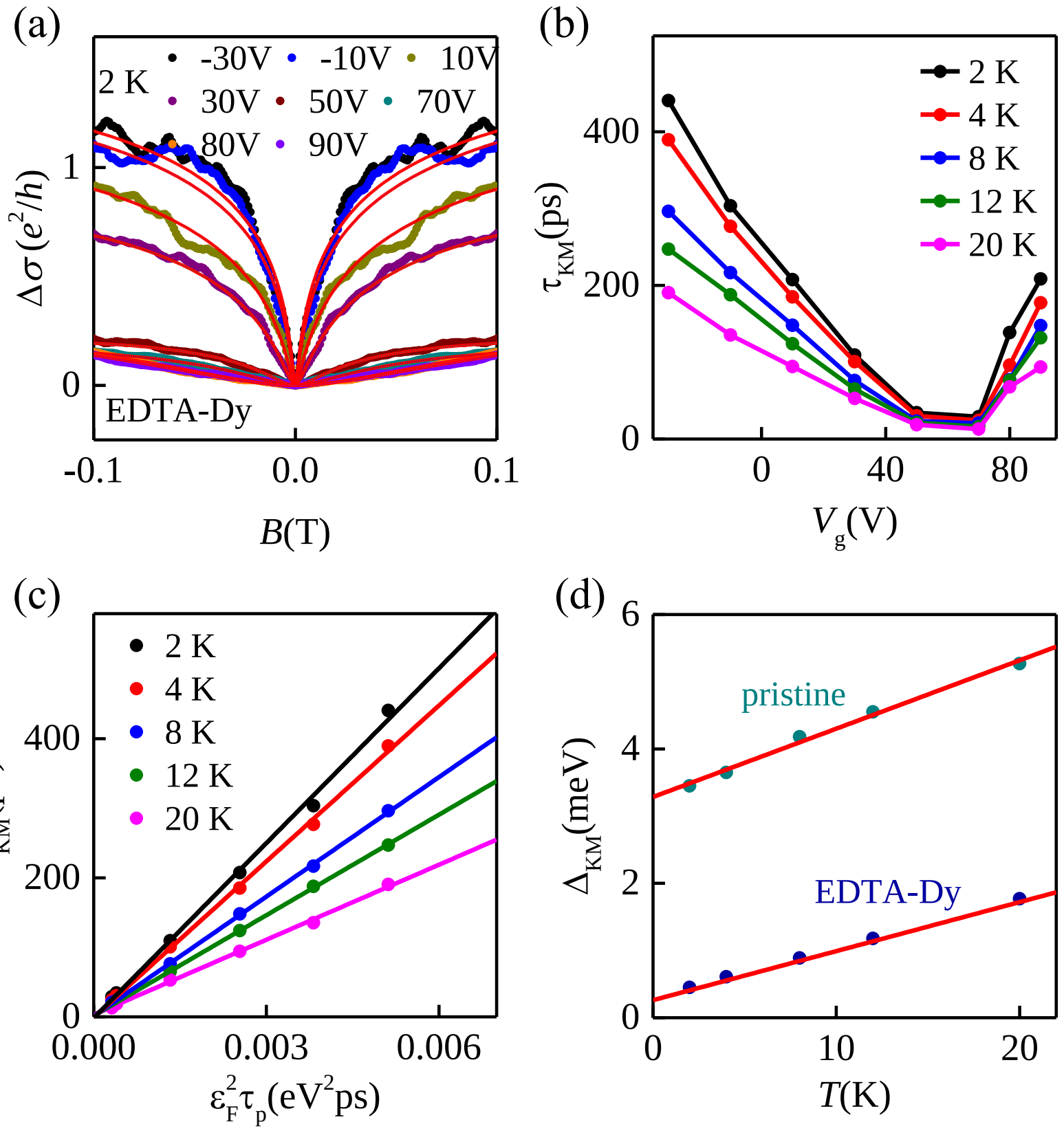

**Figure 2. Selectively enhanced KM-type SOI in graphene and Elliot-Yafet plot.** (a) The WL features of EDTA-Dy decorated graphene for different $V_g$ at 2 K. The spin scattering time can be extracted from fitting, and the red curves are the fitting curves. (b) The $V_g$ dependent spin scattering time of EDTA-Dy decorated graphene at various temperatures. (c) Fitting between $\tau_{KM}$ and $\tau_p$ using $\tau_{KM} = \tau_p(\varepsilon_F/\Delta_{KM})^2$, resulting in the SOI strength of EDTA-Dy decorated graphene. (d) The $\Delta_{D0}$ is obtained by extrapolating the temperature to zero, which is 0.26 meV for pristine graphene, and 3.3 meV for EDTA-Dy decorated graphene.

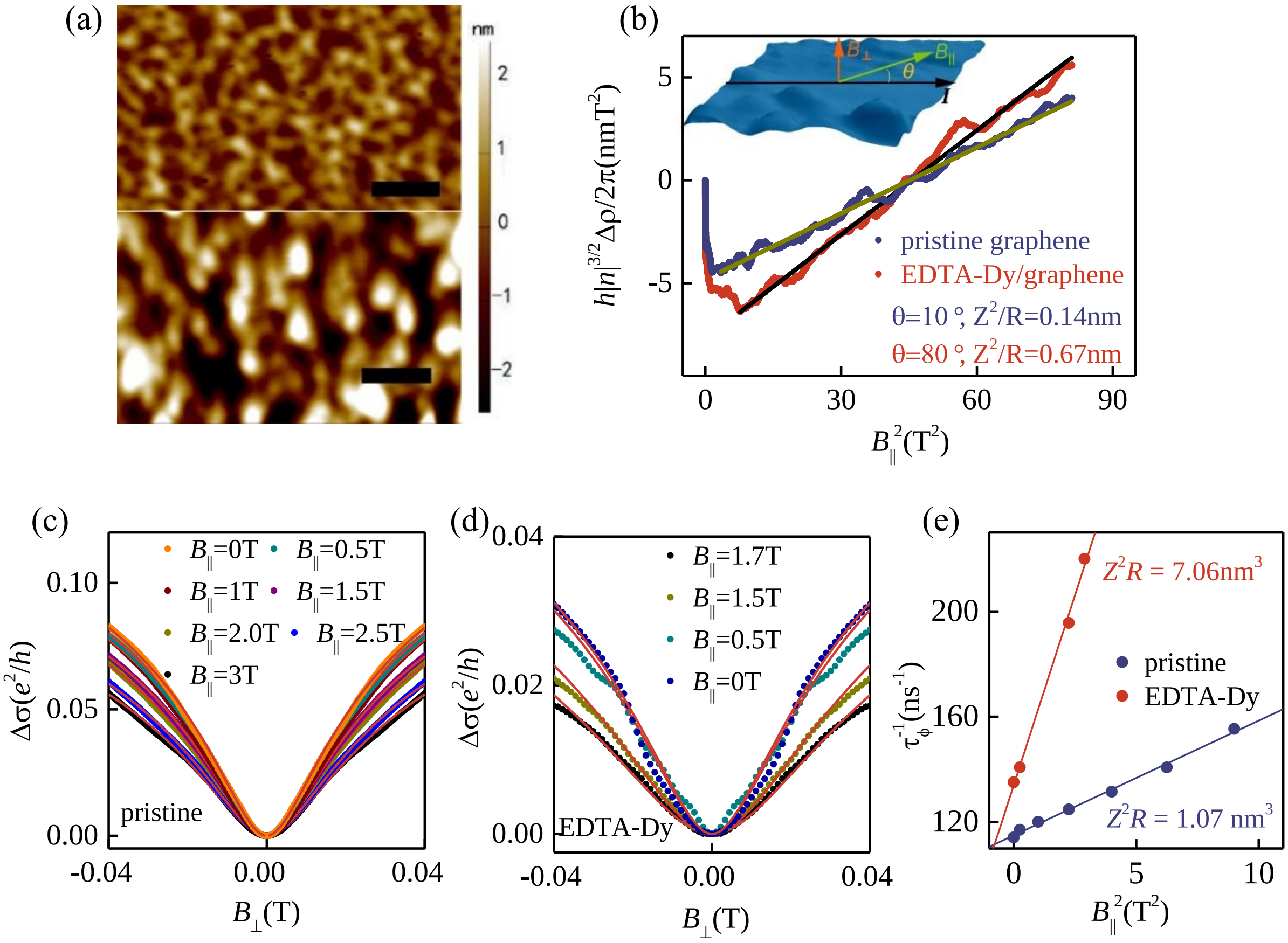

**Figure 3. Ripple configuration revealed by the vector magnet measurement.** (a) AFM images of pristine graphene (upper) and EDTA-Dy decorated graphene (down). The scale bar is 100 nm. (b) Resistivity of pristine graphene and EDTA-Dy decorated graphene dependent on $B_{\parallel}^2$. The solid lines are the fitting according to Eq. (2) using $n$ = $6.44\times10^{12}$ cm$^{-2}$ and $n = 4.27\times10^{12}$ cm$^{-2}$. The inset is the measurement configuration. (c, d) $B_{\perp}$-dependent magnetoconductivity ($B_{\perp}$<0.04 T), at a series of fixed $B_{\parallel}$, for the pristine graphene and EDTA-Dy decorated graphene respectively. The lines are the fitting according to Eq. (3). (e) Extracted values of $\tau_{\varphi}^{-1}$ plotted against $B_{\parallel}^2$. The lines are the fitting according to Eq. (4).

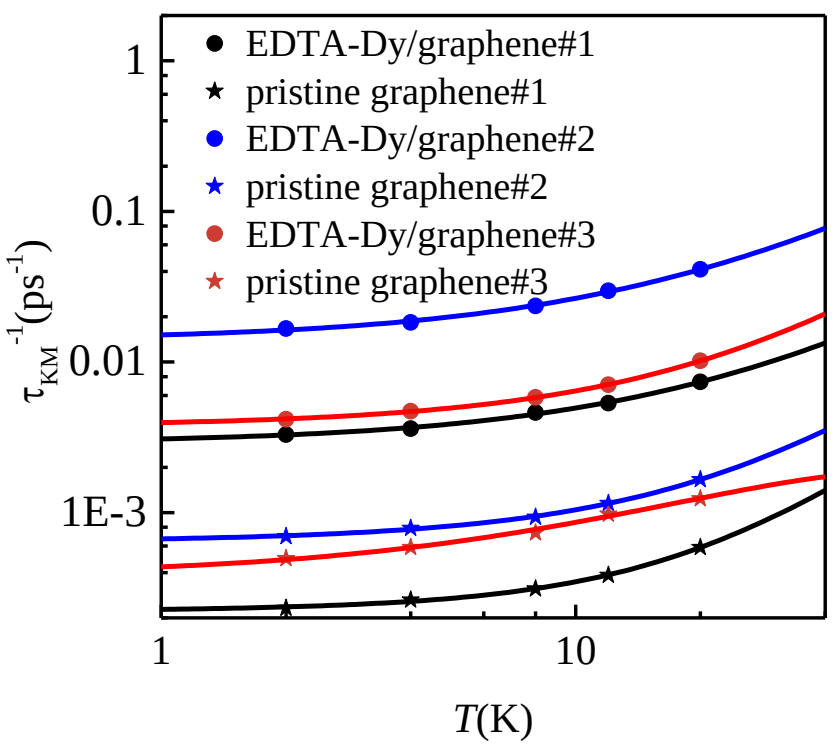

**Figure 4. Fitting the spin relaxation by the flexural phonon model.** It shows the spin life time as a function of temperatures. Circle and star dots correspond to the graphene devices before and after EDTA-Dy decoration. The solid curves are the fitting plot according to the flexural phonon theory.

| Sample | $n(10^{12}cm^{-2})$ | $\mu(cm^2\ V^{-1}\ s^{-1})$ | $l_e$(nm) | $D(cm^2s^{-1})$ | $Z$(nm) | $R$(nm) |
| --- | --- | --- | --- | --- | --- | --- |
| #1 | 10.68/4.28 | 1919/3226 | 64.86/71.67 | 324/358 | 0.62/1.48 | 2.7/3.2 |
| #2 | 9.95/2.73 | 5342/10160 | 174.3/415.7 | 871/2078 | N.A. | N.A. |
| #3 | 4.01/2.72 | 3033/6112 | 168.9/268.3 | 844/1342 | N.A. | N.A. |
| #4 | 5.29/3.09 | 1174/6161 | 31.52/112.6 | 158/563 | N.A. | N.A. |

**Table I. Device parameters for four samples before (left) and after EDTA-Dy decorated (right). N.A. means that the value is not measured.**